\documentclass[letterpaper,twocolumn,prl,aps,superscriptaddress,amsmath,amssymb,floatfix]{revtex4-1}
\usepackage{mathptmx}
\usepackage[latin9]{inputenc}
\usepackage{color}
\usepackage{amsmath}
\usepackage{amssymb}
\usepackage{graphicx}
\usepackage{esint}
\usepackage{enumitem}
\usepackage[unicode=true,
 bookmarks=true,bookmarksnumbered=false,bookmarksopen=false,
 breaklinks=false,pdfborder={0 0 1},backref=false,colorlinks=true]
 {hyperref}
\hypersetup{
 linkcolor=magenta,urlcolor=blue,citecolor=blue,pdfstartview={FitH}}

\makeatletter

\pdfpageheight\paperheight
\pdfpagewidth\paperwidth

\usepackage{textcomp}
\usepackage{epstopdf}

\pdfpageheight\paperheight
\pdfpagewidth\paperwidth

\@ifundefined{textcolor}{}{%
 \definecolor{BLACK}{gray}{0}
 \definecolor{WHITE}{gray}{1}
 \definecolor{RED}{rgb}{1,0,0}
 \definecolor{GREEN}{rgb}{0,1,0}
 \definecolor{BLUE}{rgb}{0,0,1}
 \definecolor{CYAN}{cmyk}{1,0,0,0}
 \definecolor{MAGENTA}{cmyk}{0,1,0,0}
 \definecolor{YELLOW}{cmyk}{0,0,1,0}
}

\usepackage{xcolor}
\usepackage{soul}
\definecolor{blue}{rgb}{0,0,1}
\definecolor{red}{rgb}{1,0,0}
\definecolor{green}{rgb}{0,1,0}

\usepackage{soul}

\makeatother

\begin{document}

\title{Fiber-to-chip grating couplers for Lithium Niobate on Sapphire}

\author{{Xiang~Chen}}
\affiliation{Laboratory of Quantum Information, University of Science and Technology of China, Hefei 230026, China.}
\affiliation{Anhui Province Key Laboratory of Quantum Network, University of Science and Technology of China, Hefei 230026, China}

\author{Jia-Qi Wang}
\affiliation{Laboratory of Quantum Information, University of Science and Technology of China, Hefei 230026, China.}
\affiliation{Anhui Province Key Laboratory of Quantum Network, University of Science and Technology of China, Hefei 230026, China}

\author{Yuan-Hao Yang}
\affiliation{Laboratory of Quantum Information, University of Science and Technology of China, Hefei 230026, China.}
\affiliation{Anhui Province Key Laboratory of Quantum Network, University of Science and Technology of China, Hefei 230026, China}

\author{Zheng-Xu Zhu}
\affiliation{Laboratory of Quantum Information, University of Science and Technology of China, Hefei 230026, China.}
\affiliation{Anhui Province Key Laboratory of Quantum Network, University of Science and Technology of China, Hefei 230026, China}

\author{Xin-Biao Xu}
\affiliation{Laboratory of Quantum Information, University of Science and Technology of China, Hefei 230026, China.}
\affiliation{Anhui Province Key Laboratory of Quantum Network, University of Science and Technology of China, Hefei 230026, China}
\address{CAS Center for Excellence in Quantum Information and Quantum Physics, University of Science and Technology of China, Hefei 230026, China}

\author{Ming Li}
\affiliation{Laboratory of Quantum Information, University of Science and Technology of China, Hefei 230026, China.}
\affiliation{Anhui Province Key Laboratory of Quantum Network, University of Science and Technology of China, Hefei 230026, China}
\address{CAS Center for Excellence in Quantum Information and Quantum Physics, University of Science and Technology of China, Hefei 230026, China}

\author{Xi-Feng Ren}
\affiliation{Laboratory of Quantum Information, University of Science and Technology of China, Hefei 230026, China.}
\affiliation{Anhui Province Key Laboratory of Quantum Network, University of Science and Technology of China, Hefei 230026, China}
\address{CAS Center for Excellence in Quantum Information and Quantum Physics, University of Science and Technology of China, Hefei 230026, China}

\author{Guang-Can Guo}
\affiliation{Laboratory of Quantum Information, University of Science and Technology of China, Hefei 230026, China.}
\affiliation{Anhui Province Key Laboratory of Quantum Network, University of Science and Technology of China, Hefei 230026, China}
\address{CAS Center for Excellence in Quantum Information and Quantum Physics, University of Science and Technology of China, Hefei 230026, China}

\author{Chang-Ling Zou}
\email{clzou321@ustc.edu.cn}
\affiliation{Laboratory of Quantum Information, University of Science and Technology of China, Hefei 230026, China.}
\affiliation{Anhui Province Key Laboratory of Quantum Network, University of Science and Technology of China, Hefei 230026, China}
\address{CAS Center for Excellence in Quantum Information and Quantum Physics, University of Science and Technology of China, Hefei 230026, China}

\date{\today}

\begin{abstract}
Lithium Niobate on Sapphire (LNOS) is an emerging platform for photonic integrated circuits, offering unique properties such as a wide transparency window, high nonlinearity, and strong electro-optic, nonlinear, and acousto-optic effects. Efficient light coupling between optical fibers and on-chip waveguides is crucial for practical applications. We present the design, simulation, and experimental demonstration of high-efficiency fiber-to-chip grating couplers for LNOS. The grating coupler design employs a self-imaging approach with a fixed period and linearly diminished filling factor, enabling a negative diffracted angle to match the fiber array and suppress higher-order diffraction. Numerical simulations predict a coupling efficiency of 42\% at 1550\,nm wavelength. The grating couplers are fabricated on an X-cut, 400\,nm thick LN film with a 220\,nm etching depth using electron beam lithography and inductively coupled plasma etching. Experimental characterization using a fiber array and a 6-axis displacement stage reveals a single-end coupling efficiency exceeding 20\%, confirming the effectiveness of the design. The demonstrated grating couplers pave the way for efficient light coupling in LN-on-Sapphire photonic circuits, enabling diverse applications in classical and quantum information processing, sensing, and nonlinear optics.
\end{abstract}

\maketitle

\section{Introduction}

Photonic integrated circuits (PICs) have emerged as a powerful platform for classical and quantum information processing, enabled by the flexibility of integrating various basic photonic devices on a single chip~\cite{Saleh1991, Elshaari2020, Shekhar2024}. PICs offer high stability, large scalability, low power consumption, and promising unprecedented photonic functionalities due to the greatly enhanced light-matter interaction with tightly confined optical fields. Among various material platforms, thin-film lithium niobate (LN) stands out for its outstanding optical properties~\cite{Boes2018, Zhu2021, Boes2023, Xie2024}, including a wide bandgap, strong $\chi^{(2)}$ nonlinearity, and strong electro-optic and piezoelectric effects. As a result, the LN photonic chip platform has attracted significant attention in recent years, and great progress has been achieved. LN-on-insulator (LNOI) has found excellent applications ranging from electro-optic modulators~\cite{Wang2018}, microcombs~\cite{Zhang2019}, and high-efficiency frequency converters~\cite{Xu2021, Han2021, Lu2021} to microwave photonics~\cite{Feng2024}.

While most previous studies on thin-film LN-based photonics have focused on LNOI, where LN is supported by a silicon substrate with a silica buffer layer, the LN-on-Sapphire (LNOS) platform offers unique advantages for realizing functional and hybridized photonic devices~\cite{Sarabalis2020, Mishra2021, Mayor2021, Shao2022, Celik2022, Yang2023}. LNOS overcomes the limitations of LNOI, such as the strong attenuation of acoustic waves by the silica buffer layer and the non-transparency of the silicon substrate at visible wavelengths. The single-crystal sapphire substrate of LNOS provides excellent thermal and mechanical stability. In particular, the LNOS promises the guiding and confining of GHz phonons and is compatible with superconducting circuits via piezo-mechanical couplings, making it an ideal choice for hybrid photonic-phononic-superconducting devices in quantum applications~\cite{Xu2022}. Furthermore, the wide transparency window of sapphire, extending into the visible spectrum, allows for the realization of LN-based devices operating at shorter wavelengths. This is particularly crucial for interfacing with Rydberg atom~\cite{Chen2022, Isichenko2023} arrays and trapped ions~\cite{Mehta2020, Niffenegger2020, Saha2023}, which are promising platforms for scalable quantum computation and simulation.

To harness the full potential of LNOS for these cutting-edge applications, efficient light coupling between optical fibers and on-chip waveguides is critical. Various approaches for fiber-to-chip coupling have been proposed and demonstrated. Near-field couplers based on tapered nanofibers have been explored to realize ultra-high efficiency and broadband coupling~\cite{Iecke2015, Sulway2023}. Still, they require precise near-field manipulation and are not extensible to fiber arrays. Edge coupling, which relies on mode matching between the optical fiber and the waveguide facet, can provide high efficiency and broadband operation but requires precise alignment and polishing of the waveguide facets, which can be time-consuming and costly. Grating couplers offer a convenient method and have been widely applied due to their low cost, high robustness, and compatibility with standard fiber array packaging techniques~\cite{Taillaert2004, Ding2014}. Recently, highly efficient self-imaging apodized grating couplers have been proposed and demonstrated for the LNOI platform~\cite{Lomonte:21}, meriting its photonic applications for their easy manufacturability and high coupling efficiency of 47.1\% at telecom wavelengths. However, the design and realization of grating couplers specifically tailored for the LNOS platform have remained largely unexplored~\cite{Sarabalis2020}. 

This work bridges this gap by presenting the design, simulation, and experimental demonstration of high-efficiency fiber-to-chip grating couplers for LNOS. By employing the self-imaging approach and optimizing the grating parameters, we achieve a coupling efficiency of up to 42\% at 1550\,nm in our simulations, using a fixed period and linearly diminished filling factor. Our experimental results validate these designs, demonstrating a coupling efficiency of 20\% and a bandwidth exceeding 20\,nm. The LNOS platform, with its unique combination of high-quality phononic waveguides and compatibility with superconducting qubits, opens new avenues for groundbreaking hybrid photonic chip platforms. Our work provides the essential chip-to-fiber interconnects to accelerate the development of photonic-atom hybrid chips and hybrid phononic-photonic-superconducting quantum circuits. These advances hold immense promise for realizing scalable and functional hybrid quantum devices that leverage the strengths of various quantum systems and pave the way toward chip-based quantum information processors and sensors.

\section{Structures and Principle}

\begin{figure}[h!t]
    \centering
    \includegraphics[width=\linewidth]{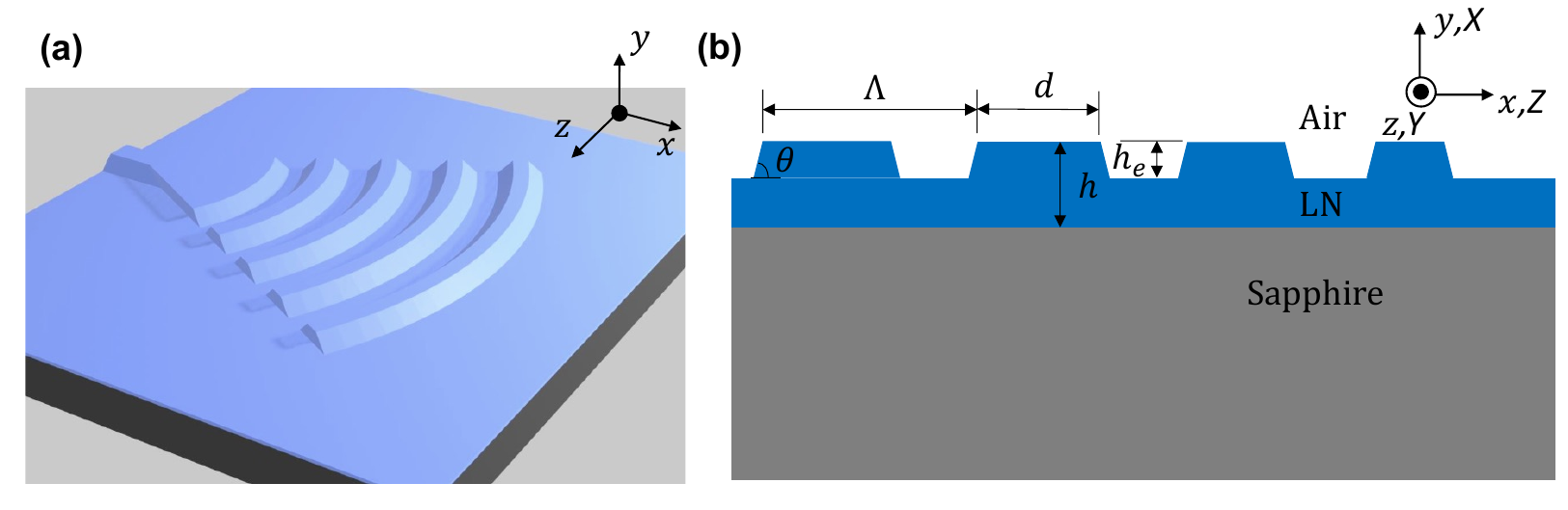}
    \caption{(a) The schematic illustration of the grating coupling structure on a LNOS platform. (b) The x-y cross-sectional view of the grating coupler with the geometry parameters labels, also as the schematic of our two-dimensional model. The axis written in uppercase is the intrinsic coordinate of LN, and the one in lowercase is the model coordinate. }
    \label{fig1}
\end{figure}

Figure~\ref{fig1}(a) illustrates the proposed grating coupler structure for the LNOS platform, which is designed without dielectric cladding on the top and without a buffer layer or a metal layer on the bottom of the grating coupler. The device is designed for X-cut LN thin films.
The three-dimensional (3D) grating structure, as shown in Fig.~\ref{fig1}(a), is designed according to the self-imaging method that was introduced by Ref.~\cite{Lomonte:21}. The grating coupler is mirror-symmetric and consists of two parts: a straight waveguide and a sector with etched concentric circles. The detailed geometric structure is depicted in the two-dimensional (2D) cross-section in Fig.~\ref{fig1}(b), along with the symmetric axis of the device. The blue and grey regions represent the LN and sapphire layers, respectively. The key parameter of the grating coupler is the filling factor (FF), which is defined as the ratio of the unetched region width ($d$) to the grating period ($\Lambda$), i.e., $\textrm{FF} = d / \Lambda$.

We aim to couple the light in the waveguide into the fiber in the free space. What we are going to couple here is the fundamental mode in the waveguide, TE$_0$ mode.
The operating principle of the grating coupler relies on the etched grooves, which introduce a periodic perturbation to the waveguide and allow light to be scattered from the guided mode to free space and subsequently collected by an optical fiber, as well as from the guided mode to the substrate. The scattering process is governed by the phase-matching condition, also known as the Bragg equation:
\begin{equation}
    \frac{2\pi}{\lambda} n_c \sin \theta_c = \frac{2\pi}{\lambda} n_\mathrm{eff} - q \frac{2\pi}{\Lambda}
    \quad (q = 1,2,\dots)
    \label{Eq1}
\end{equation}
where $\lambda$ is the wavelength of light, $n_c$ represents the refractive index of the cover medium (air in this case), $\theta_c$ is the diffraction angle in the air, $n_\mathrm{eff}$ is the effective refractive index of the guided mode in the grating region, and $q$ is the diffraction order. The $n_\mathrm{eff}$ can be calculated as the weighted average of unetched and etched regions because the approximately linear relationship between $n_\mathrm{eff}$ and the thickness of LN works well in this case.

Because there might exist different solutions for different $q$, guided light could be diffracted at different angles. However, a fiber end-face can only match one of the diffracted beams. Therefore, we design the grating to have a negative first-order diffraction angle ($\theta_c < 0$), ensuring that there is only one solution of $\theta_c$ for $q = 1$ and the diffraction for higher orders is suppressed. It is also important to note that the diffraction of light into the substrate follows Eq.~(\ref{Eq1}) as well. Consequently, for each allowed diffraction beam with order $q$ and angle $\theta_c$ in the cover medium, there exists a corresponding solution of diffraction angle $\theta_s$ at the same order in the substrate, indicating that energy loss to the substrate is inherent to the coupling process. This situation significantly differs from that of LNOI, where an optimal buffer layer thickness can help suppress the diffraction into the substrate via destructive interference.

In the following sections, we will discuss optimizing the grating coupler design using numerical simulations and present the experimental results demonstrating the high coupling efficiency achieved with our proposed structure.



\section{Designs}

To determine the optical set of geometry parameters for the grating coupler, we performed finite-domain time-difference (FDTD) simulations. Since the numerical simulation of the full 3D model is time-consuming, we began with 2D simulations to roughly determine the parameters and performances of the device. Then, we use the full 3D model to characterize the device.

\subsection{Two-Dimensional Simulation}

\begin{figure*}[htbp]
    \centering
    \includegraphics[width=0.75\linewidth]{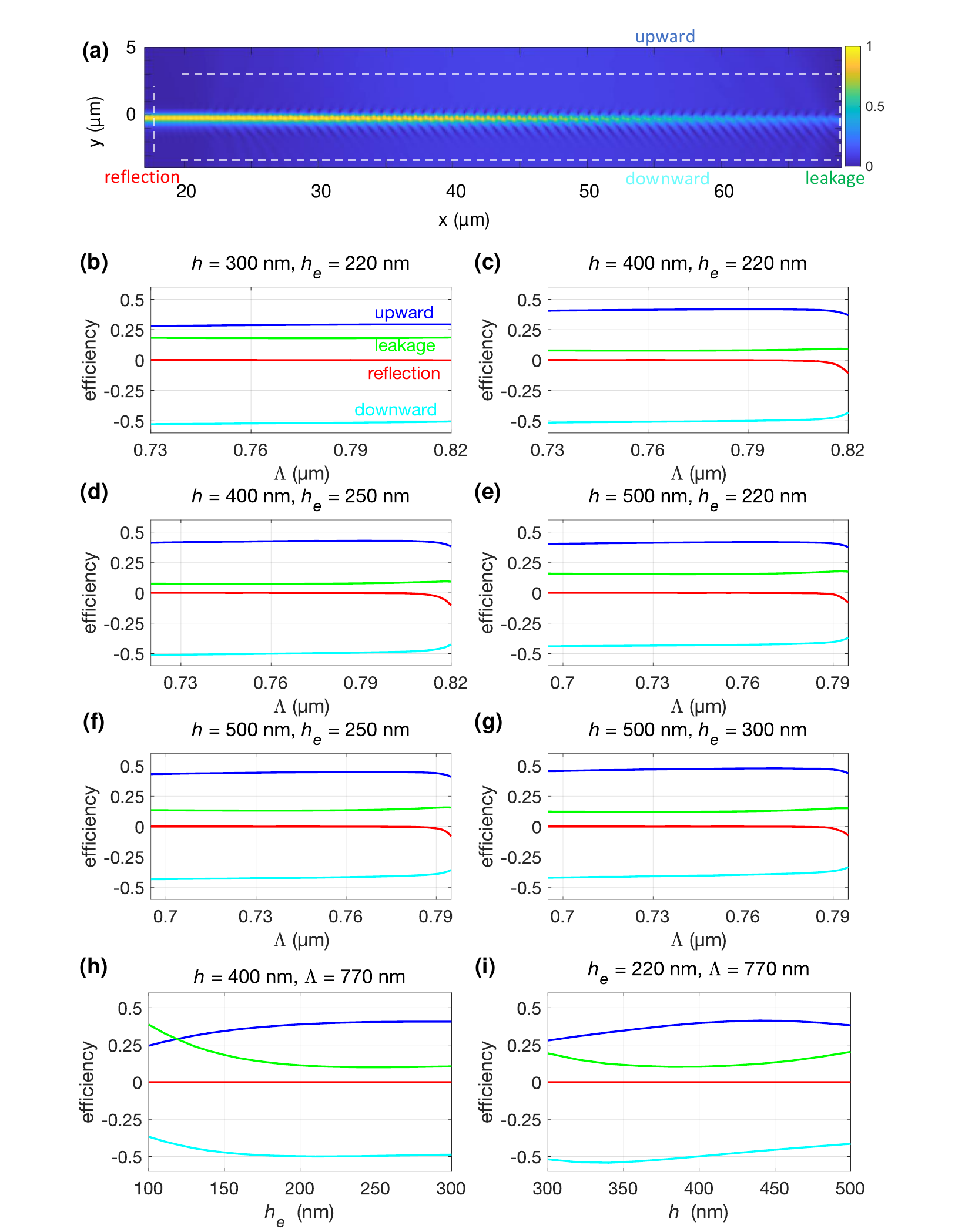}
    \caption{(a) Field distribution at the near field when light coupling out from the waveguide. The four white dashed lines represent the four detectors we set. (b)-(i) Parameter dependences (thickness of LN $h$, etching depth $h_e$ and period $\Lambda$), where blue curves, yellow curves, purple curves, and red curves represent energy coupling upwards, transmit, reflected, and downwards respectively. {Here, the efficiency values incorporate directional information based on energy flux direction, with positive values indicating upward diffraction and transmission, and negative values indicating reflection and substrate leakage.} }
    \label{fig2}
\end{figure*}

Due to the symmetry of the device, the simplified 2D model is performed on the structure shown in Fig.~\ref{fig1}(b). For a simple waveguide structure, the 2D model could provide a relatively precise estimation of the performances because the effective refractive index $n_\mathrm{eff}$ difference between in a 2D thin film waveguide model and the 3\,$\mu m$-wide waveguide is less than 1\%. As a first step, we solved for the eigenmodes of the waveguide and used the fundamental mode (TE$_0$ mode) as the input light source. We then set up four monitors to detect the power of the upward, downward, transmitted, and reflected diffracted light, the four white dashed lines shown in Fig.~\ref{fig2}(a). This approach allows us to efficiently estimate the energy coupling out of the chip in the x-y plane and optimize the grating coupler design by neglecting the degree of freedom along the z-direction.

{Because there is no reflection from the substrate, we set the thickness of the sapphire as $5\,\mathrm{\mu m}$. The number of grating teeth was fixed at $N=60$, and the FF was set to decrease from 85\% to 25\% along the waveguide linearly.} Initially, we focused on maximizing the upward diffracted energy, enabling near-field simulations to quickly identify a suitable parameter range. Figure~\ref{fig2}(a) shows a typical simulation result, illustrating the decay of the guided optical field as it is diffracted upward and downwards when passing through the grating region.

Through a series of parameter scans, we identified the LN film thickness $h$, etching depth $h_e$, and grating period $\Lambda$ as the most critical parameters influencing the coupling efficiency. Figures~\ref{fig2}(b)-(g) present a selection of simulation results for various parameter sets. These results reveal that, for a fixed LN thickness $h$, increasing the etch depth $h_e$ leads to higher diffraction efficiency. This is because deeper etching corresponds to a stronger perturbation of the mode in the waveguide, resulting in a stronger waveguide mode to free space beam coupling. When the film becomes thicker, the effective refractive index $n_\mathrm{eff}$ increases, and so a smaller grating period $\Lambda$ is required to maintain optimal coupling. Notably, we found that a larger $\Lambda$ leads to a die-off in efficiency because the mode in the waveguide tends to couple into the backward direction TE$_0$ mode.
Furthermore, if the film is too thin, a larger fraction of the diffracted energy from the mode will be coupled into the substrate, thus limiting the efficiency to couple light upward. Conversely, if LN is too thick, the etched grooves have fewer perturbations in the guiding mode. However, due to the restriction in etching LN, e.g. 220\,nm in our experiments. Therefore, achieving a relatively high perturbation of the mode will be challenging, and thus the diffraction efficiency by the grating is limited.
Considering various factors discussed above, we choose $h=400$\,nm, $h_e=220$\,nm for fixed parameters of the LNOS platform, and $\Lambda=770$\,nm as the rough estimation of optimal grating parameter for further investigations by the 3D model.


\begin{figure*}[h!t]
    \centering
    \includegraphics[width=0.75\linewidth]{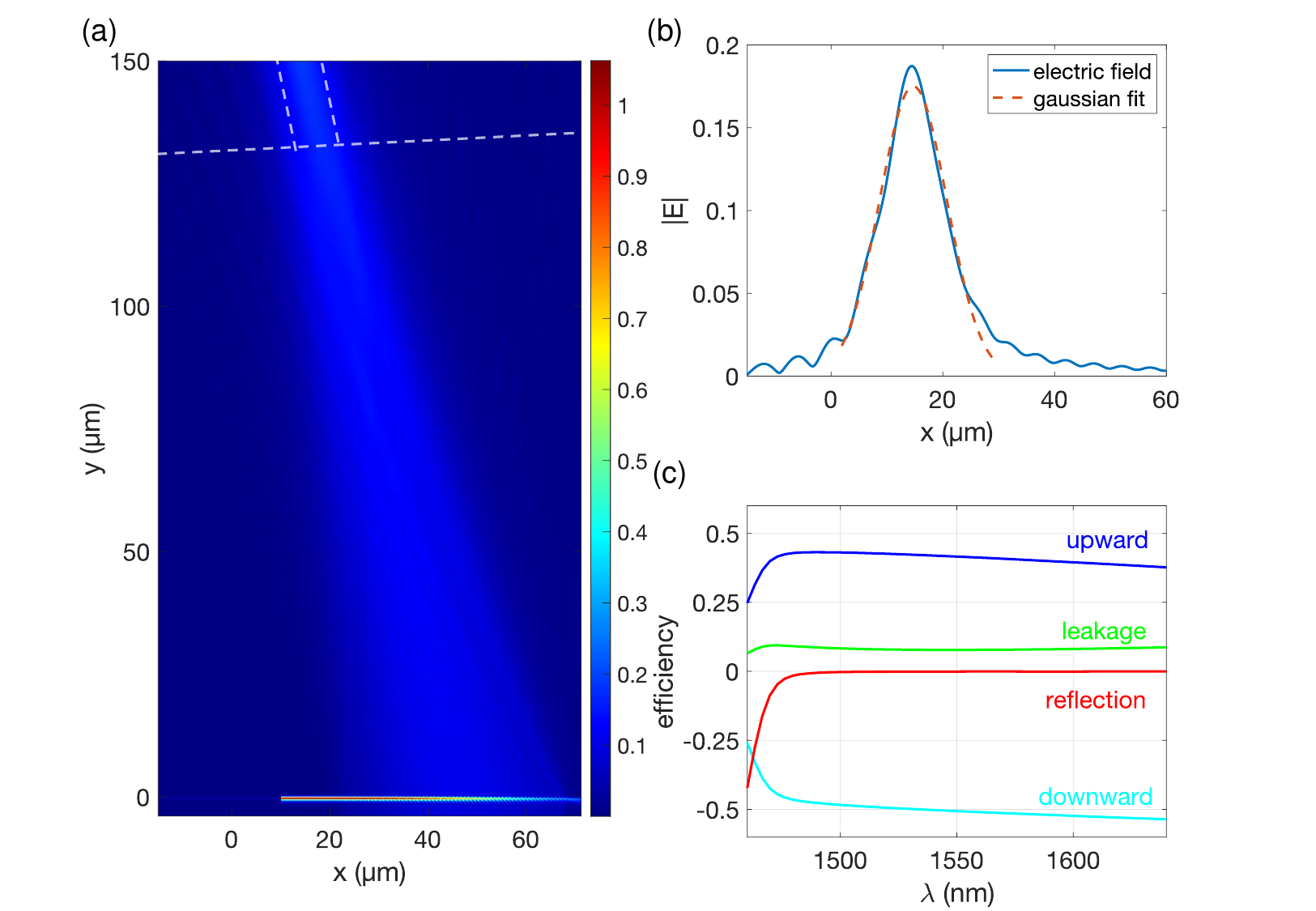}
    \caption{(a) Field distribution at the far field when 1550\,nm light coupling into the waveguide. The white dashed line represents the approximate position of the fiber. (b) Far-field electric field cross-section (at $y=150\,\mathrm{\mu m}$), fitted by Gaussian distribution ($\sigma = 5.03\,\mathrm{\mu m}$). (c) The wavelength dependences for infrared wavelength.}
    \label{fig3}
\end{figure*}

For the goal of achieving optimal coupling with a fiber end-face, the matching between the beam profile and the fiber mode profile is also important. Thus, we performed far-field simulations to evaluate the propagation of the diffracted light in free space. In our design, we apply a linear varying FF from 85\% to 25\%, and the resulting far-field distribution was numerically solved and shown in Fig.~\ref{fig3}(a). Thanks to the self-imaging design, the diffracted light forms a well-shaped focused beam as the diffracted light focuses at a height. By adjusting the changing rate of FF, we can control the focusing intensity and the location of the focal point.

In Fig.~\ref{fig3}(a), the spot size is minimized at a height of about 150\,$\mu m$ above the chip surface. We placed a monitor at this location to record the energy flow, and the results are shown in Fig.~\ref{fig3}(b). The beam profile is well-fitted by a Gaussian distribution. By fitting the data with a Gaussian distribution, we determined that the mode field diameter (MFD) of the diffracted beam is 12.05\,$\mu m$, the potential excellent matching between the beam and the fiber mode profile, at least along the x-axis, because the mode profile of a standard single-mode fiber (SMF-28, 8-deg polished, 250\,$\mu m$ bare, FC/APC, Thorlabs) can also be approximated by a Gaussian profile with a MFD of 10.4\,$\mu m$.

Figure~\ref{fig3}(c) shows the wavelength dependencies performance of the grating coupler for the optimized set of parameters. The curves exhibit a relatively flat response, confirming that the grating coupler is broadband. Notably, upward efficiency remains above 40\% across a wide wavelength range from 1500\,nm to 1600\,nm. This broadband performance is highly desirable for many applications, as it enables efficient coupling of light over a wide range of wavelengths without the need for significant adjustments to the grating coupler design.

{As a result, we have chosen $h = 400$\,nm, $h_e = 220$\,nm, $\Lambda = 770$\,nm, and FF linearly diminishing from 85\% to 25\% as our optimized set of parameters in the following sections. }

\subsection{Three-Dimensional Simulation}

While 2D simulations provide a valuable reference for the grating coupler design, they cannot fully capture the behavior of diffracted light due to the absence of performance along the z-axis. This limitation may lead to significant discrepancies between the simulated and experimental results, particularly in terms of the shape and quality of the diffracted beam. To address this issue and obtain a more accurate representation of the grating coupler performance, we conducted 3D-FDTD simulations. {In our 3D simulations, we incorporated the full anisotropic permittivity tensor for X-cut LN, ensuring accurate modeling of the mode propagation and diffraction characteristics.} The waveguide width was set to 3\,$\mu m$, and the apodized grating coupler's open angle was 60° to ensure that most of the energy (over 99\%) is confined within the sector region. However, due to the limitation in computational memory, it is challenging to simulate the entire structure directly. Instead, we adopted a two-step simulation approach. First, we simulated a thin layer of the grating coupler near the chip surface using 3D-FDTD to obtain the near-field distribution of the diffracted light. Then, we used the Rayleigh-Sommerfeld (R-S) diffraction integral to propagate the near-field distribution to the far-field region. 

\begin{figure*}[h!t]
    \centering
    \includegraphics[width=0.75\linewidth]{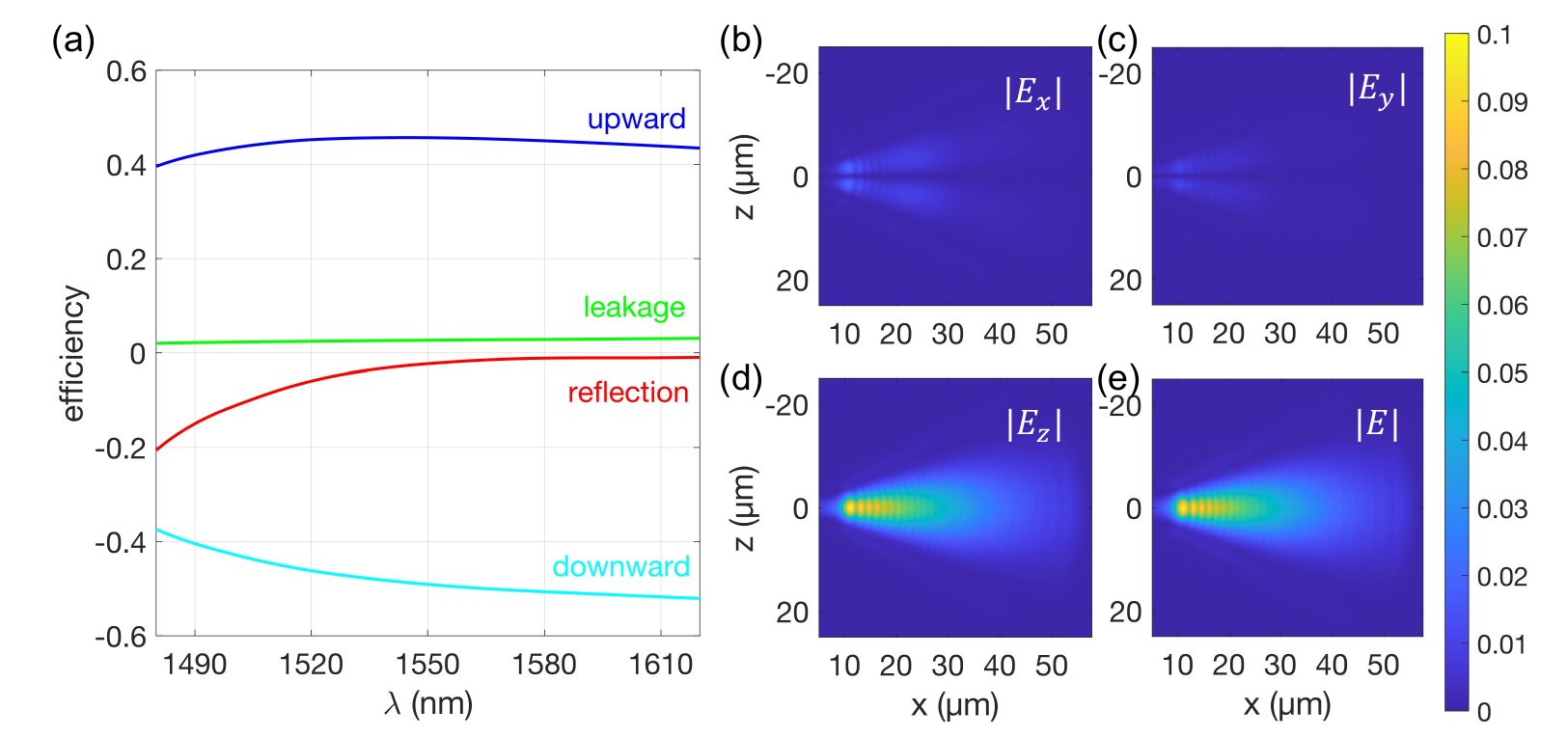}
    \caption{The near-field 3D full-vectorial simulation results. (a) The wavelength dependences. (b)-(e) Compare with the three upward diffraction electrical vector components. Four figures show $|E_x|$, $|E_y|$, $|E_z|$, and $|E|$ respectively, at the height of $y=3\,\mathrm{\mu m}$. }
    \label{fig5}
\end{figure*}

In the first step, we set up monitors to record the power flux spectra in the upward, downward, transmitted, and reflected directions, similar to the 2D simulations. As shown in Fig.~\ref{fig5}(a), the wavelength dependence of the coupling efficiency remained relatively weak, confirming the results of the 2D model.

We extracted the electric field distribution from the top monitor in the 3D-FDTD simulations for the far-field beam properties. We extracted the electric field distribution from the top monitor in the 3D-FDTD simulation. Figures~\ref{fig5}(b)-(e) show the three components of the electric field ($|E_x|$, $|E_y|$, and $|E_z|$) and the total electric field magnitude ($|E|$). Among the three components, $|E_z|$ is the most prominent, which is consistent with the fact that the guided mode in the waveguide is predominantly TE-polarized (TE$_0$).

Then, according to Green's function approach in solving Maxwell's Equations, the far-field properties of the diffracted light can be determined by the fields at a boundary.
However, if we accurately calculate the full vectorial of the electric field, it will be equivalent to FDTD. Here, we apply the scalar refraction approximation and use a scalar field to represent the vector field. This approximation can be justified based on the factor that $|E_z|$ is the most prominent component in our simulations. So, we can apply the R-S diffraction integral to calculate the far-field distribution. We should note that since we only extract information from a finite area, some diffraction light does not pass through this port, resulting in certain truncation errors. That is not an issue because we are only concerned with the shape of the spot.

\begin{figure}[ht]
    \centering
    \includegraphics[width=\linewidth]{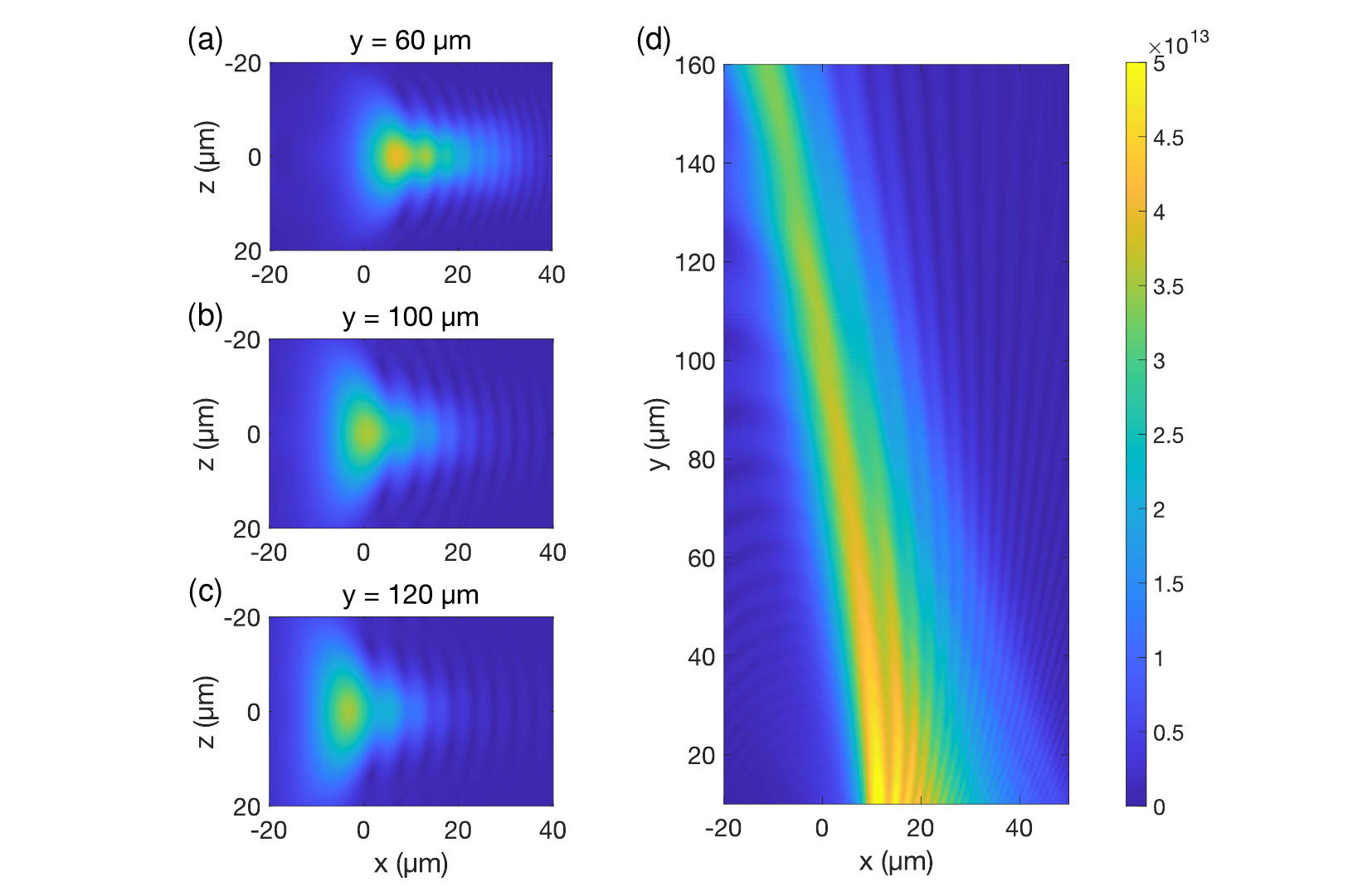}
    \caption{Three-dimensional field distribution at far-field when applying R-S diffraction. (a)-(c) Two-dimensional cross-sections in X-Z planes, respectively at $y= 60\,\mathrm{\mu m}$, $100\,\mathrm{\mu m}$, and $120\,\mathrm{\mu m}$. (d) Two-dimensional cross-section of diffraction light in X-Y plane. }
    \label{fig6}
\end{figure}

Then, We obtained a series of cross-sectional plots of the far-field distribution at different heights from the chip surface. Figures~\ref{fig6}(a)-(c) shows three representative cross-sections at heights of $y= 60\,\mathrm{\mu m}$, $100\,\mathrm{\mu m}$, and $120\,\mathrm{\mu m}$.
These results reveal how the diffracted light evolves as it propagates away from the chip. Interestingly, we found that the beam profile at the height of $y=100\,\mathrm{\mu m}$ {exhibits a concentrated distribution in an elliptical-like region}, indicating good focusing in both the x and z directions. However, at the height of $y=120\,\mathrm{\mu m}$, the focusing effect in the x-direction is more pronounced than in the z-direction, suggesting that further optimization of the grating coupler design may be necessary to achieve a symmetrical focal spot. The asymmetry in the focusing behavior along the x and z directions can be attributed to the different convergence rates in these two directions, which is influenced by the FF distribution along the grating. While this issue could potentially be addressed by fine-tuning the FF distribution, such an optimization process would be time-consuming and computationally expensive using 3D-FDTD simulations. Therefore, we opted to defer this optimization to the experimental stage, where the grating parameters can be more efficiently refined based on the measured coupling efficiency and beam profile.

To validate the accuracy of the R-S diffraction integral approach, we compared the calculated far-field distribution in the X-Y plane with the results obtained from the 2D model. As shown in Fig.~\ref{fig6}(d), the far-field patterns obtained from both methods are in good agreement, confirming the validity of our approaches. The presence of ripples in the far-field pattern can be attributed to the truncation of the near-field data.

\section{Experimental results}

\begin{figure*}
    \includegraphics[width=0.75\textwidth]{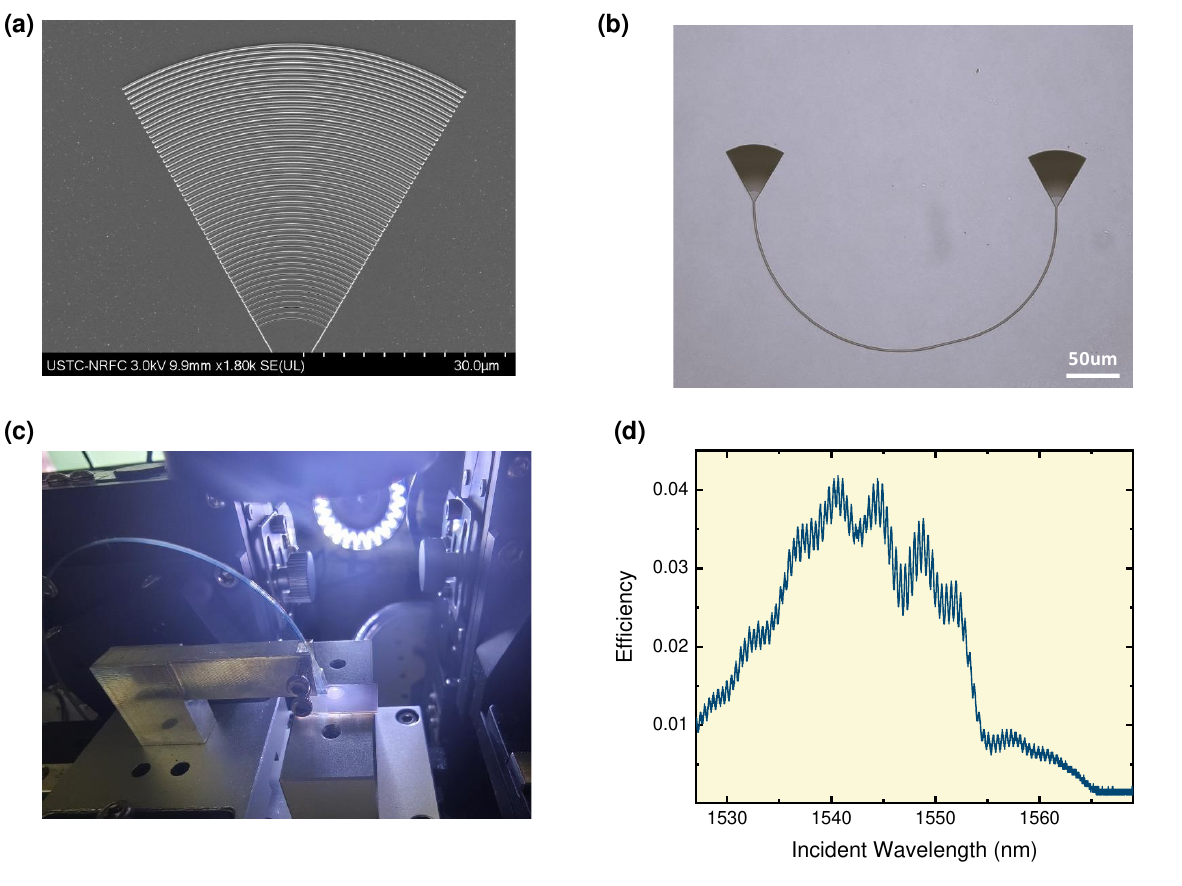}
    \caption{Grating devices and characterization. (a) The SEM image of the fabricated LNOS grating coupler showing uniform eteching profile and smooth sidewalls. (b) Optical microscope image of the device for characterizing the transmission properties of grating, with input and output grating connected by a waveguide. (c) Experimental setup featuring fiber array coupling, tunable lasers source, and alignment stage for precise positioning. (d) Measured transmission spectrum of the devices shown in (b).}
    \label{fig7}
\end{figure*}


According to the optimized parameters from 3D simulation, we fabricated the grating coupling devices for LNOS, with the X-cut LN thickness is $400\,$nm and the etching depth is $220\,$nm. The fabrication process began with the deposition of Hydrogen silsesquioxane (HSQ) layer on the top of LN using spin coating. The grating coupler patterns were then exposed using electron beam lithography (EBL), and subsequently transferred to the LN layer using inductively coupled plasma (ICP). After a LN etching depth of 220$\,$nm is achieved, the residual photoresist is removed using buffered oxide etchant (BOE). Figure~\ref{fig7}(a) shows a scanning electron microscope (SEM) image of the fabricated grating coupler, revealing {an open angle of $60^\circ$ with smooth and uniform etching profile of the LN layer. The fabricated grating features a period of 770\,nm with filling factor varying from 80\% to 20\%, resulting in feature sizes ranging from 154\,nm (narrowest etched grooves) to 616\,nm (widest etched grooves).}


To characterize the efficiency of the fabricated grating couplers [Fig.~\ref{fig7}(b)], We employed a fiber array for coupling from the vertical direction, as illustrated in Fig.~\ref{fig7}(c). The fiber array was mounted on a 6-axis displacement stage, which allows precise positioning along  X-Y-Z axes and corresponding angular adjustment for optimizing the coupling efficiency. Light from a tunable telecom laser was coupled into the waveguide on the chip through the input grating coupler, and the transmitted light through the waveguide is collected from the output grating coupler.

By scanning the wavelength of the laser, the transmission spectrum is recorded, as shown in Fig.~\ref{fig7}(d). According to the figure, we found that the highest transmission efficiency exceeds 4\% at around 1545\,nm. {Since the spectrum is measured for light transmitted through two grating couplers, the corresponding efficiency of single grating coupler is estimated as $\sqrt{4\%}\approx20\%$, which has proven to be significantly beneficial for our subsequent experiments. The corresponding 3~dB bandwidth of a single grating coupler exceeds 25\,nm, which corresponds to the total transmission through both couplers remains above $(20\%/2)^2\approx1\%$. }

The experimental results of the efficiency and the bandwidth are both smaller than the numerical results.
{The measured bandwidth is mainly limited by the wavelength-dependent diffraction angle for the output beam from the grating (according to Eq.~(\ref{Eq1})). Since the location of the collection fiber remained fixed during the measurements, the coupling efficiency reduces by changing the wavelength as the beam deviates from optimal alignment with the fiber core.}
We attributed the degradation of efficiency to two imperfections: (1) due to the limited fabrication precision, the fabricated device might deviated from the optimized structure geometry, and surface roughness or defects of the grating structures could introduces extra optical losses. {Cross-sectional SEM analysis revealed a sidewall etching angle of about 55°, which our simulations confirm has negligible impact on the coupling efficiency.} (2) the alignment mismatches between the fiber and the diffracted beam. The plane of the fiber end-face might not at the focus plane of the diffracted beam, and the alignment precision is also limited in practice. 

\section{Discussions}

The successful demonstration of high-efficiency grating couplers on the LNOS platform opens up various avenues for future research and development. This section discusses the potential studies and applications that can be pursued based on our grating coupler design.

First, the integration of our high-efficiency grating couplers with other hybrid photonic components. While this work focuses on the design and characterization of the grating coupler itself, future efforts should be directed towards the development of a complete hybrid photonic integrated circuits by leveraging the unique properties of LNOS system. For example, the Brillouin scattering between guided phonon and photon in the waveguide on the LNOS platform have been demonstrated with end-face couplers~\cite{Yang2023}. The integration of our grating couplers with these Brillouin-active waveguide is helpful in further investigations of microwave-to-optical frequency conversion, microwave photonics, Brillouin lasers and gyroscopes, acoustic-optics modulators~\cite{Qin2024}, as well as non-recioprocal devices.

Second, the optical and microwave co-packaging of LNOS chips. The compatibility of the grating coupler design with fiber array packaging is a significant advantage, as it enables reliable and stable coupling. When combined with wirebonding technique for coupling to microwave ports, the packaging of the LNOS chip becomes crucial for their deployment in real-world scenarios. Furthermore, the long-term stability and reliability of the packaged LNOS under various environmental conditions, particularly at sub-Kelvin cryogenic temperatures, is of great importance for potential hybridization with superconducting qubits~\cite{Xu2022}.

Third, the extension to other wavelengths. While our work primarily focuses on the telecom wavelength range, the design principles can be readily extended to other wavelengths, particularly in the visible spectrum. Our preliminary simulation results demonstrate the feasibility of designing grating couplers for visible wavelengths by adjusting the grating period to account for the shift in the effective refractive index. For example, by setting the grating period to $\Lambda = 390\,$\,nm, we can achieve efficient coupling at $780\,$nm, as the effective refractive index only shifts slightly. However, it is important to note that the presence of higher-order modes in the waveguide at these wavelengths may require careful numerical optimization to ensure optimal coupling efficiency and suppress any undesired modal interactions. Future research could focus on the detailed design and experimental demonstration of high-efficiency grating couplers for specific wavelengths, such as $780\,$nm that matches the D1 transition of Rubidium atoms~\cite{Liu2022a,Liu2022b,Xu2023}. This would open up exciting opportunities for chip-integrated quantum information processing with single atom array.


\section{Conclusions}

In conclusion, we have demonstrated high-efficiency fiber-to-chip grating couplers for LNOS, paving the way for practical applications of this emerging photonic platform. The self-imaging design with optimized parameters achieves a single-end coupling efficiency exceeding 20\%, with the potential for further improvement through advanced optimization techniques and the incorporation of reflective layers~\cite{Wang2024}. Our work highlights the promise of LNOS for hybrid quantum chips, which might play important roles in classical and quantum information processing and sensing.

\section*{Acknowledgments}

This work was funded by the National Key R\&D Program (Grant No.~2021YFF0603701), the Innovation Program for Quantum Science and Technology (Grant No. 2021ZD0300203), the National Natural Science Foundation of China (Grants No.~92265210, No.~12104441, 12061131011, U21A6006, 12293053), the Fundamental Research Funds for the Central Universities, and USTC Research Funds of the Double First-Class Initiative. The numerical calculations in this paper have been done on the supercomputing system in the Supercomputing Center of University of Science and Technology of China. This work was partially carried out at the USTC Center for Micro and Nanoscale Research and Fabrication.

\nocite{*}
\bibliographystyle{zou}
\bibliography{sample}


\end{document}